\documentclass[a4paper,twoside,11pt]{article}
\usepackage[utf8]{inputenc}

\usepackage[english]{babel}
\usepackage{subeqnarray}
\usepackage{url}

\usepackage{graphicx,psfrag}
\usepackage{fixltx2e,amsmath}
\usepackage{xcolor}
\usepackage{amssymb}
\usepackage[final]{pdfpages}
\usepackage{bbold}
\usepackage{dsfont}
\usepackage{pgfplots}
\usepackage{subfig}
\usepackage{algorithm}
\usepackage{algorithmic}
\usepackage[font=small,labelfont=bf]{caption}

\usepackage[font={footnotesize}]{caption}

\usepackage[utf8]{inputenc}
\usepackage[T1]{fontenc}
\usepackage{xcolor,amsmath,amssymb}
\usepackage{geometry}
\usepackage{graphicx}
\usepackage[sort&compress,numbers]{natbib}
\usepackage{enumerate}
\usepackage{wasysym}
\usepackage[textsize=footnotesize]{todonotes}
\usepackage{url}

\usepackage{pgfplots}

\definecolor{darkblue}{rgb}{0,0,0.6}
\definecolor{darkred}{rgb}{0.6,0,0}

\usepackage{hyperref}

\usepackage{authblk}

\usepackage[bitstream-charter]{mathdesign}

\usetikzlibrary{pgfplots.groupplots}

\graphicspath{{figures/}} 

\title{Did lockdowns serve their purpose?}
\author[1]{Serena Bradde}
\author[2]{Benedetta Cerruti}
\author[3]{Jean-Philippe Bouchaud }
\affil[1]{American Physical Society, Ridge, NY,  \& David Rittenhouse Laboratories, University of Pennsylvania, Philadelphia}
\affil[2]{benedetta.cerruti@gmail.com}
\affil[3]{Capital Fund Management \& Acad\'emie des Sciences, Paris}
\date{June 2020}

\begin{document}

\maketitle

\begin{abstract}
    We show that the dynamics of the number of deaths due to Covid in different countries is to a large extent universal once the origin of time is chosen to be the start of the lockdown, and the number of death is rescaled by the total number of deaths after the lockdown, itself a proxy of the number of infections at the start of the lockdown. Such a curve collapse is much less convincing when normalizing by the total population. Sweden, with its no-lockdown, light-touch approach, is the only outlier that deviates considerably from the average behavior. We argue that these model-free findings provide strong support for the effectiveness of the lockdowns in mitigating the lethality of the virus.    
\end{abstract}

\section{Introduction}

What was the real effect of lockdown measures in controlling the number of deaths in western countries? Since the economic cost of the lockdown are expected to be very high, it is an important question that sparks intense -- and politically loaded -- debates. Some commentators have suggested that the lockdowns were in fact totally useless. The dynamics of the pandemic, left on its own device, would have led to the same decrease of cases.  

In this short note, we provide some model-free evidence that the lockdown measures played a direct role in the evolution of the number of deaths. It is of course difficult to prove causality in absence of a control experiment. However, as suggested in our previous posts [1], the collapse of rescaled plots and the correlation of the total number of deaths with respect to the delay of implementing the lockdown provide a smoking gun of the beneficial effect of lockdown measures.

\section{Results}

\begin{figure}[h]
  \centering
a)\includegraphics[width=.45\textwidth]{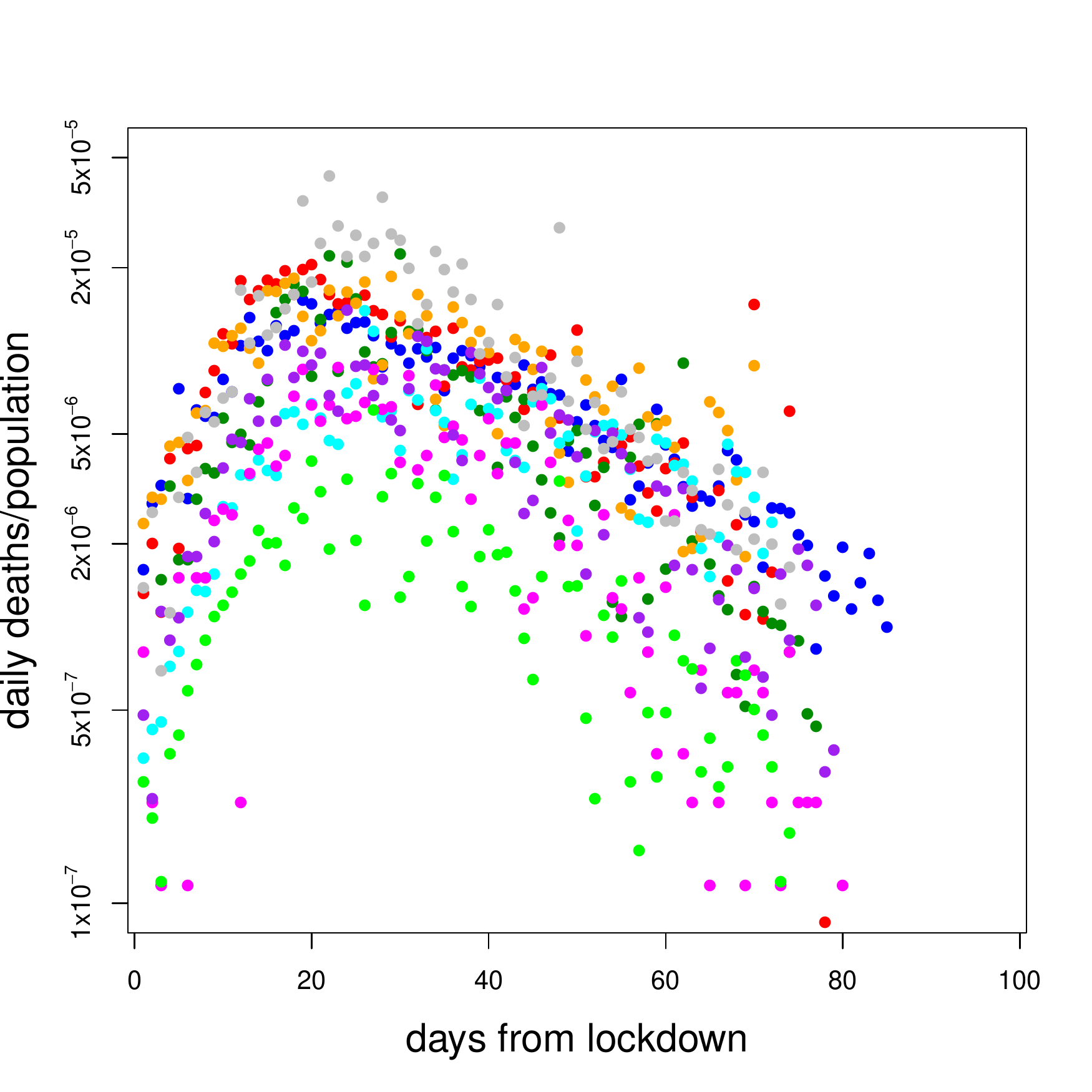}
b)\includegraphics[width=.45\textwidth]{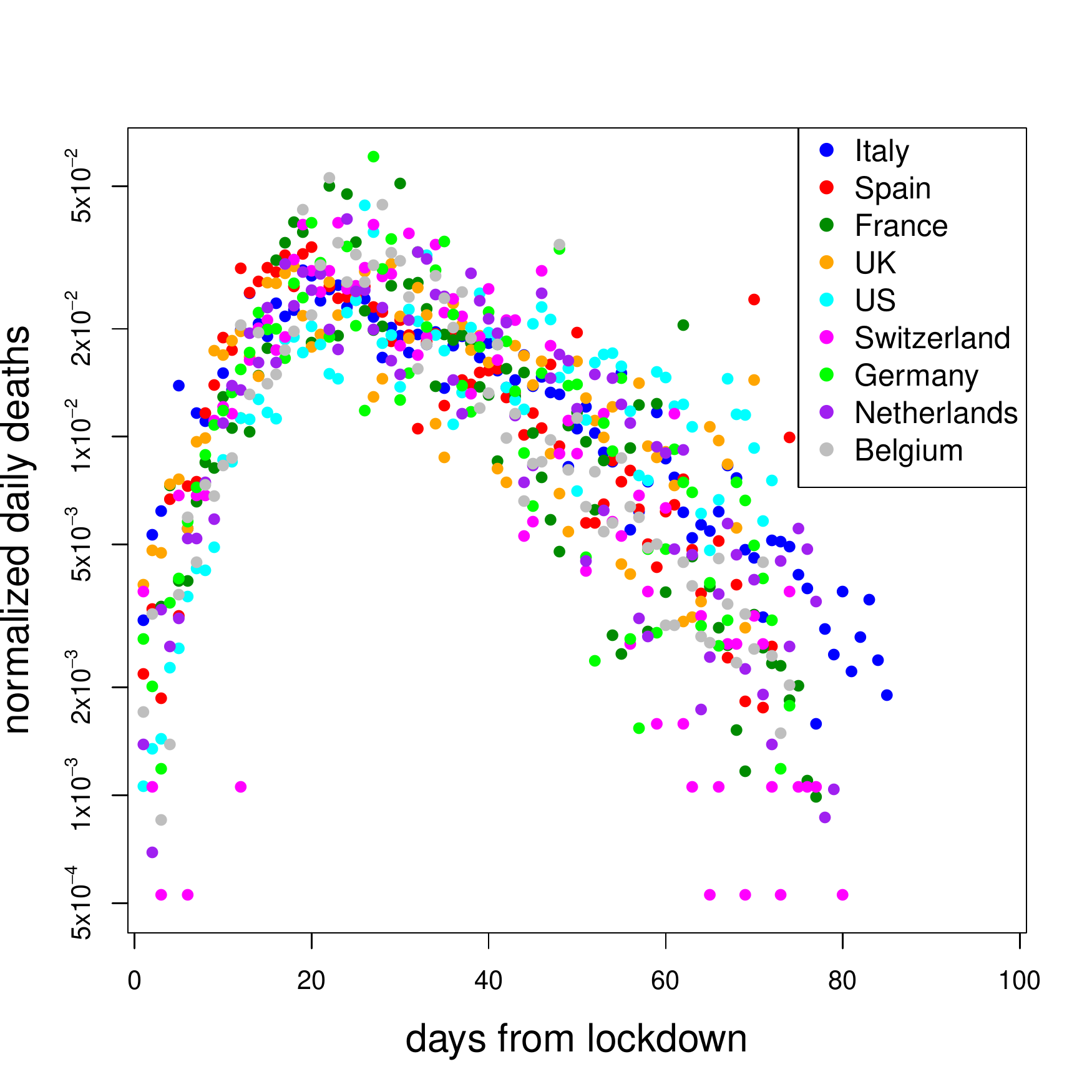}
  \caption{These plots show the number of daily deaths in each country time-shifted such that the first day of the lockdown coincides with the origin and normalized by a) the population size of each country and b) the total number of deaths up to the latest available time.} 
  \label{Fig1}
\end{figure}

In Figure 1a) we plot the daily deaths normalized by the country’s total population. The time shift is such that the time origin coincides with the starting of the restriction measures in each country, $t_{\rm{lock}}$. We already see a strong hint of the effect of the lockdown: all curves peak roughly 20 days after $t_{\rm{lock}}$, independently of {\it when} the lockdown was actually implemented. Since the number of cases at the moment of the lockdown (i.e. the development stage of the epidemic) varies from country to country, this is already quite remarkable.\footnote{For another study emphasizing the universality of the dynamics across countries, see [10].} 

Furthermore, the plot shows that Germany has one of the lowest death rates per inhabitants while Belgium one of the highest. If the reproduction rate of the virus was still above $1$ after the beginning of the lockdown, it would have eventually reached the whole population. Since one expects a similar fraction of susceptible people, roughly the same fraction of the population should have died in all countries. In other words, we would in that case expect a collapse of all the curves for different European countries when normalized by the total population size. 

Instead, Fig. 1a)  shows that countries that responded faster, had a much smaller impact in terms of daily deaths contradicting our initial assumption of a freely circulating virus. 

Hence, we investigate the opposite extreme assumption: the lockdown did completely stop the circulation of the virus, meaning that the virus would not affect the whole population but only individuals who were already infected. If this was true, the collapse of the curves should happen when normalized by the number of infected at the time of lockdown ($t_{\rm{lock}}$) instead of the total population. Since we do not have the real number of infected at $t_{\rm{lock}}$, we use a proxy given by the total number of deaths from lockdown $t_{\rm{lock}}$ up to the latest available time $t^\star$. Indeed, the total number of deaths in perfect isolation, is proportional to the number of infected at the first day of lockdown. Figure 1b) shows the collapse of the curves of daily deaths time-shifted and normalized by the total number of deaths up to the latest available time.

We can see by eye that the collapse in Figure 1b) is better than the one in Figure 1a), in line with our assumption that only infected people at the start of the lockdown contributed to the subsequent death rate. In order to quantify the quality of the collapse, we plot the average of all the curves (after a 10-days moving average) normalized by population size (red) and by total number of deaths (black) and their dispersion in Figure 2. The dispersion around the peak is found to be a factor 2.2 smaller for the curves normalized as in Figure 1b).

\begin{figure} [h]
  \centering
\includegraphics[width=.7\textwidth]{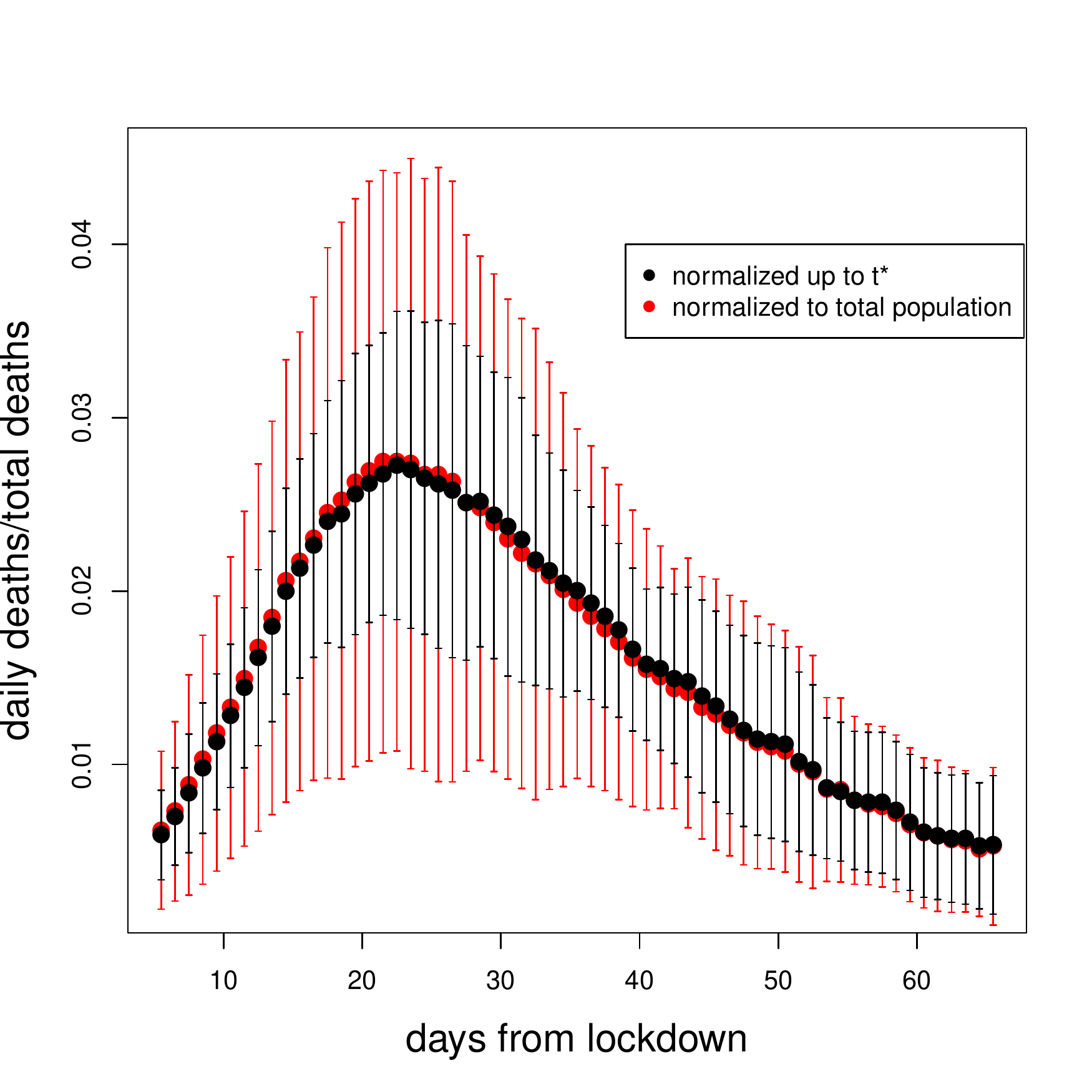}
  \caption{The dots and the error bars show the average and the dispersion of the 10-days moving average curves shown in Figure 1a) (red) and in Figure 1b) (black). Since the two average curves have different y-ranges, we fix their integral to be one.} 
  \label{Fig2}
\end{figure}

We could argue that in reality the lockdown was not perfect and differed from country to country: this could justify the variability when the number of deaths becomes smaller. This simple argument suggests that lockdown measures, when implemented faster, reduce the number of infected sharply after lockdown and consequently the number of deaths. 

In addition, there is another argument suggesting that the lockdown controlled the number of deaths. As mentioned above, all the curves show a peak at $t_{\rm{peak}}\sim$20 days after $t_{\rm{lock}}$. This value is also equal to typical time from infection to death, ${\bar T}_{\rm{lag}}$. Thus, most of the individuals that contributed to the daily deaths were infected ${\bar T}_{\rm{lag}}$ days before the peak, exactly at $t_{\rm{lock}}$. We want to stress that the distribution of time from infection to death, $\rho(T_{\rm{lag}})$, is log-normal, i.e. very broad and not sharply peaked as was assumed by the author of Ref. [2]. This means that the results presented in [2] are misleading. If the distribution of time from infection to death, $\rho(T_{\rm{lag}})$, was sharply peaked around ${\bar T}_{\rm{lag}}$ days with a variance $\sigma$, we could have learned the death growth rate (called $\gamma$ in [2]) in absence of lockdown measures simply by fitting the curves from patient-0 day to time $t_{\rm{lock}}+{\bar T}_{\rm{lag}}-\sigma$. But in fact $\rho(T_{\rm{lag}})$ has fat tails: people start dying much before the average lag time. This means that the effect of lockdown on the daily deaths curves was already important from day 1 to ${\bar T}_{\rm{lag}}$ after $t_{\rm{lock}}$, while this effect is completely disregarded in the analysis of Ref. [2].

Lockdown has became a familiar term that refers to a set of rules put in place in different countries to reduce people's mobility and keep interactions to a minimum in order to avoid the spreading of the virus. Unraveling the net effect of lockdown with respect to less extreme measures like social distancing and the demand to wear masks is not easy. Still, since Sweden never enforced strict lockdown but only light-touch measures, we can compare the Swedish national data with the rest of European countries (see Figure 3). The plot shows how the functional form of the curves is very different. We can identify two strong features that are specific to Sweden: first, the peak of the Sweden curve (blue curve) is delayed by $\sim 10$ days and does not coincide with ${\bar T}_{\rm{lag}}$; second the Sweden curve has a fat tail, that decays much slower than in other countries. This suggests that, if European countries had followed Sweden's example, they would have paid a much higher cost in terms of human lives.
 
\begin{figure} [h]
  \centering
\includegraphics[width=.7\textwidth]{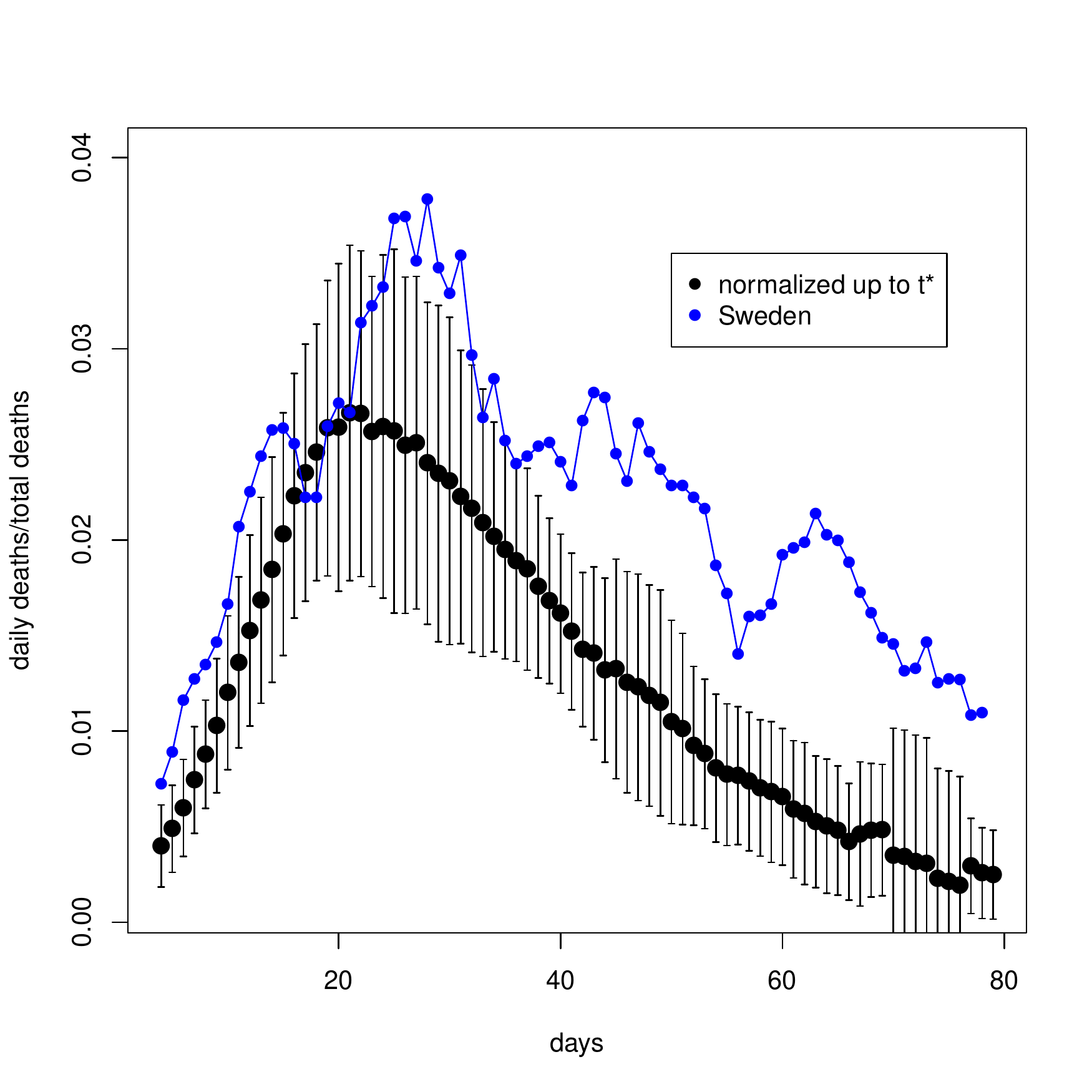}
  \caption{The dots and the error bars show the average and the dispersion of a 7-days moving average curves, similarly to what shown in Figure 1b) (black), compared to the Swedish data (blue). We chose here day 0 for Sweden as the day when the number of deaths in Sweden were about the same of the number of deaths in Germany at $t_{lock}$ ($\sim$ 20), and normalize so that the average curve maximum, at day=18, coincides with Swedish day 18 ordinate. The Sweden curve is outside the dispersion range.} 
  \label{Fig2}
\end{figure}

\section{Conclusions}

Due to their high economic cost, there has been a heated debate on the contribution of lockdown measures on the spreading of the epidemic. Arguing that the effect of the lockdown could not be visible until the average delay from rising of the symptoms to deaths, some papers, like [2], even claimed that in absence of any lockdown measures the number of deaths would have been less than actually observed. As shown in [3], this is not the case. The distribution of time delay from infection to death is very broad (log-normal) and positively skewed, so we expect that more than half of the people would have died before the average lag time. So, it is misleading to fit the data just after lockdown to estimate what would happen without implementing the lockdown.

In this note, we showed evidence of the effect of lockdown on the curve of daily deaths. Firstly, we see that all curves peak $\sim$20 days after $t_{\rm{lock}}$: this corresponds to the number of days from infection to death when we expect more than half of the infected at $t_{\rm{lock}}$ to have died already. After $t_{\rm{lock}}$, the number of new infections becomes negligible so the majority of people dying were those infected before the lockdown was implemented. 

Sweden, the only European country in this analysis that did not implement any lockdown, shows a delayed peak and a different dynamical pattern (Figure 3), confirming our initial hypothesis that the epidemic spreading stopped quite abruptly because of the lockdown.

A possible alternative explanation is that the virus had in fact reached all susceptible people in each country.  But in this case, the total deaths should have been correlated with the size of each nation, not to the number of infected people at the start of the lockdown. However, rescaling by population size leads to a dispersion 2.2 times larger at the peak than rescaling by the total number of deaths (Figure 2), suggesting that the saturation hypothesis is not warranted, in line with several previous studies as well [4,5,6]. A similar idea was also discussed in Ref. [4] where the authors provide further evidence i.e. serology studies and the cumulative per-capita daily deaths, to reach to same conclusion. 

Even if correlation does not prove causality, the quality of the curve collapse shown in Figure 1b) and the complementary analysis proposed by the authors in [7,8,9] provide strong evidence of a positive effect of lockdown measures in curbing the number of fatalities.

\vskip 1cm
We thank F. Zamponi, M. Kolanovic, E. Marinari, F. Valka and E. Jani for useful discussions.
\newpage
\section*{References}

[1] https://medium.com/@benedetta.cerruti/the-lockdown-general-and-special-effects-57a74ee74888
https://medium.com/@benedetta.cerruti/critical-response-time-e66cf1907cde

\noindent
[2] https://www.medrxiv.org/content/10.1101/2020.04.24.20078717v1.full.pdf

\noindent
[3] Flaxman et al. DOI: https://doi.org/10.25561/77731

\noindent
[4] https://onlinelibrary.wiley.com/doi/10.1002/jmv.25908

\noindent
[5] Bendavid, E., et al., COVID-19 Antibody Seroprevalence in Santa Clara County, California. medRxiv, 2020: p. 2020.04.14.20062463.

\noindent
[6] Salje, H., et al., Estimating the burden of SARS-CoV-2 in France. 2020.

\noindent
[7] https://www.thelancet.com/journals/lancet/article/
PIIS0140-6736(20)31357-X/

\noindent
[8] https://medium.com/@enzo.marinari/a-concise-remark-about-the-apparent-growing-lethality-of-covid-19-and-about-the-lockdown-effects-cc4a14bcb48c

\noindent
[9] https://medium.com/@tomaspueyo/\\coronavirus-should-we-aim-for-herd-immunity-like-sweden-b1de3348e88b

\noindent
[10] M. Radiom, J.-F. Berret, Common trends in the epidemic of Covid-19 disease, EPJ+, to appear

\end{document}